# Pressure-dependent structure of BaZrO$_3$ crystals as determined by Raman Spectroscopy

Dong-Hyeon Gim [1], Yeahan Sur [1], Yoon Han Lee [1], Jeong Hyuk Lee [1], Soonjae Moon [2], Yoon Seok Oh[3,*], and Kee Hoon Kim[1,4,*]

[1] *Center for Novel States of Complex Materials Research, Department of Physics and Astronomy, Seoul National University, Seoul 08826, Republic of Korea*
*loyard@snu.ac.kr (D.-H.G.); yhsur2011@gmail.com (Y.S.); lyh9042@gmail.com (Y.H.L.); jeonghyuklee1508@gmail.com (J.H.L.)*

[2] *Department of Physics, Hanyang University, Seoul, 04763, Republic of Korea*
*soonjmoon@hanyang.ac.kr*

[3] *Department of Physics, Ulsan National Institute of Science and Technology, Ulsan 44919, Republic of Korea*

[4] *Institute of Applied Physics, Department of Physics and Astronomy, Seoul National University, Seoul 08826, Republic of Korea*

\* Correspondence: ysoh@unist.ac.kr (Y.S.O.); khkim@phya.snu.ac.kr (K.H.K.)

**Abstract:** The structure of dielectric perovskite BaZrO$_3$, long known to be cubic at room temperature without any structural phase transition with variation of temperature, has been recently disputed to have different ground state structures with lower symmetries involving octahedra rotation. The pressure-dependent Raman scattering measurements can identify the hierarchy of energetically-adjacent polymorphs, helping in turn understand its ground state structure at atmospheric pressure. Here, Raman scattering spectra of high-quality BaZrO$_3$ single crystals grown by the optical floating zone method are investigated in a pressure range from 1 atm to 42 GPa. First, based on the analyses of the infrared and Raman spectra measured at the atmosphere, it is found that all observed vibrational modes can be assigned according to the cubic $Pm\bar{3}m$ structure. In addition, by applying pressure, two structural phase transitions are found at 8.4 and 19.2 GPa, one from the cubic to the rhombohedral $R\bar{3}c$ phase and the other from the rhombohedral to the tetragonal $I4/mcm$ phase. Based on the two pressure-induced structural phase transitions, the true ground state structure of BaZrO$_3$ at room temperature and ambient pressure is corroborated to be cubic while the rhombohedral phase is the closest second.

## 1. Introduction

The dielectric oxide BaZrO$_3$ (BZO) with the perovskite structure exhibits high structural stability, low thermal conductivity, and good refractory characters so that it has been widely used as e.g. thermal barrier coatings in aerospace industries [1] and as inert crucibles [2] and substrates [3] in the laboratories. With its intrinsic dielectric constant as high as 43 [4], BZO shows enhanced piezoelectric properties upon being alloyed with BaTiO$_3$ [5]; Ba(Zr,Ti)O$_3$ has been thus widely used in lead-free piezoelectric actuators, transducers, sensors, and for wireless communications [6]. The application of BZO extends to development of proton conductors [7], hydrogen separation reactors [8,9], and humidity sensors [10]. Befitting such a variety of applications requiring thermal and structural stability, BZO is known to maintain its cubic structure from 2 to 1473 K according to neutron and X-ray diffraction (XRD) studies [11,12].

However, several density functional theory (DFT) calculations have proposed that the true structural ground state be the tetragonal *I*4/*mcm* symmetry due to unstable antiphase antiferrodistortive (AFD) phonons of oxygen octahedra [13–15], and that the local tetragonal distortions be averaged-out and become undetectable in the diffraction experiments which identify the cubic $Pm\bar{3}m$ BZO structure. Seemingly supporting the DFT calculation results, a bunch of peaks were observed in the Raman spectra of BZO [16–19] although an ideal cubic perovskite has no first-order Raman-active optical phonons [20]. Hence, the observed Raman modes were interpreted to represent the short-ranged structural distortions in BZO [17–19]. Besides, in addition to the octahedral rotations, a Brillouin light scattering experiment claimed the loss of centro-symmetry due to local distortion in a commercially available BZO crystal in a temperature range from 93 to 1273 K [21]. Concrete interpretation on the vibrational modes observed in one of highest-quality BZO single crystals is therefore necessary to resolve whether a symmetry breaking is intrinsic or not.

There was another viewpoint which ascribed the observed Raman shifts to classical two-phonon modes from the cubic structure [16], as similarly known in other cubic perovskites such as $SrTiO_3$ and $KTaO_3$ [22, 23]. Vibrational spectra measured by Raman scattering [16] and phonon density of states from inelastic neutron scattering studies [24] indeed exhibited no sign of structural phase transitions in BZO in a wide temperature range (4–1200 K), suggesting that the cubic structure remains persistently. Absence of experimental evidence of any structural phase transition in BZO was attributed to zero-point fluctuations [11] and nonlocal exchange-correlation effects [25]. As an alternative scenario, it was also suggested that the energetic proximity of *I*4/*mcm*, *Imma*, and $R\bar{3}c$ structures each involving AFD distortions [25, 26] can quench the manifestation of a certain distorted structure in BZO [27]. As the small energy differences among those tilt polymorphs can be effectively discriminated by means of external perturbation [26], application of high pressure might be an effective way to sort out the energetic hierarchy of those competing polymorphs.

A previous study of pressure-dependent Raman scattering with BZO ceramics grown by the solid-state reaction method found two structural phase transitions at 9 and 23 GPa [19], which were assigned as a transition from cubic to rhombohedral $R\bar{3}c$ phase and one from rhombohedral to orthorhombic *Imma* structure, respectively. On the other hand, another structural study on a commercial BZO powder with a synchrotron XRD measurement identified only one pressure-induced transition at 17.2 GPa from the cubic to tetragonal *I*4/*mcm* phase [28]. As the BZO ceramics grown by solid state reaction can often possess local symmetry breaking by octahedra rotation due to local strain at grain boundaries [6,29,30], the discrepancies in the number of the phase transitions and the transition pressures reported by the two former high-pressure studies might originate from polycrystalline nature of the samples. Therefore, a high-pressure experiment on a single crystal of BZO is desirable to resolve the discrepancy in the high-pressure results of the polycrystalline BZO specimens.

In this article, the Raman spectra of the BZO single crystal grown by the optical floating zone method have been investigated. Based on the comparison with infrared (IR) spectra and DFT prediction on the phonon spectra, it is shown that all the Raman shifts of the BZO crystals can be successfully assigned to the two-phonon modes in the cubic phase. At higher pressures, there exist emergent first order Raman modes, of which frequencies can be successfully assigned by the new crystal symmetry stabilized above 8.4 or 19.2 GPa. Based on the full assignment of the measured Raman spectra, it is concluded that $BaZrO_3$ crystals undergo phase transitions from the cubic to the rhombohedral $R\bar{3}c$ phase at 8.4 GPa, and subsequently to the tetragonal $I4/mcm$ phase at 19.2 GPa.

## 2. Experimental Methods

*2.1. Sample Preparation*

High quality BaZrO$_3$ single crystals were grown by the optical floating-zone technique [31] in O$_2$/Ar mixed gas environment. Polycrystalline BaZrO$_3$ feed rods were prepared as stoichiometric BaO and ZrO$_2$ were mixed, ground, pelletized, and sintered at 1650 °C for 24 hours in the air. The as-grown single crystals were annealed at 1650 °C in O$_2$ flow. The grown single crystal has been cut into the circular disk with a typical diameter of 4 mm and thickness of 1 mm. Figure 1 shows XRD results of the polished (001) surface of the BaZrO$_3$ single crystal.

*2.2. Pressure-dependent Raman spectroscopy experiment*

Back-scattered Raman spectra of the BZO single crystal were measured with a commercial Raman spectrometer (Nanobase, XperRam200™) equipped with a 20× objective lens and a Nd:YAG laser with 532 nm wavelength. To load the specimen into the diamond anvil cell (DAC) for the high-pressure Raman scattering experiments, the rectangular shaped crystal was further cut into a small piece with a typical lateral size ~ 50 μm and thickness ~ 5 μm, a widest surface of which corresponds to the (001) plane. The pressure media made of the liquid blend of methanol-ethanol with a 4:1 ratio was used. Applied pressure was estimated from the R1 photoluminescent line of ruby particles inserted inside a gasket next to the sample [32]. All the measurements were performed at room temperature (298 K), and linearly polarized laser with the power of 0.6 mW was focused as a few μm² beam spot on the diamond anvils. A polarizing filter was used to collect the parallel-polarized Raman spectra from the DAC, which are denoted by $z(xx)\bar{z}$ in figures. Here $x$ and $z$ in the Porto's notation are parallel to the $[100]_{pc}$ and $[001]_{pc}$ axes of the pseudo-cubic (pc) lattice, respectively. The polarization of the incident laser light and outgoing scattering light, denoted by ($xx$) in this case, was maintained during the high-pressure measurements. Then, the DAC was decompressed to 1 GPa and compressed subsequently to collect the unpolarized Raman spectra without using the polarizing filter, which are denoted by $z(xx+xy)\bar{z}$ in figures. The parallel- and cross-polarized spectra at atmospheric pressure [Figs. 2(a),(b)] were measured outside the DAC using the polarizer filter. As a reference, the Raman spectra were also measured in a BaZrO$_3$ polycrystalline pellet at atmosphere synthesized by the solid-state reaction method.

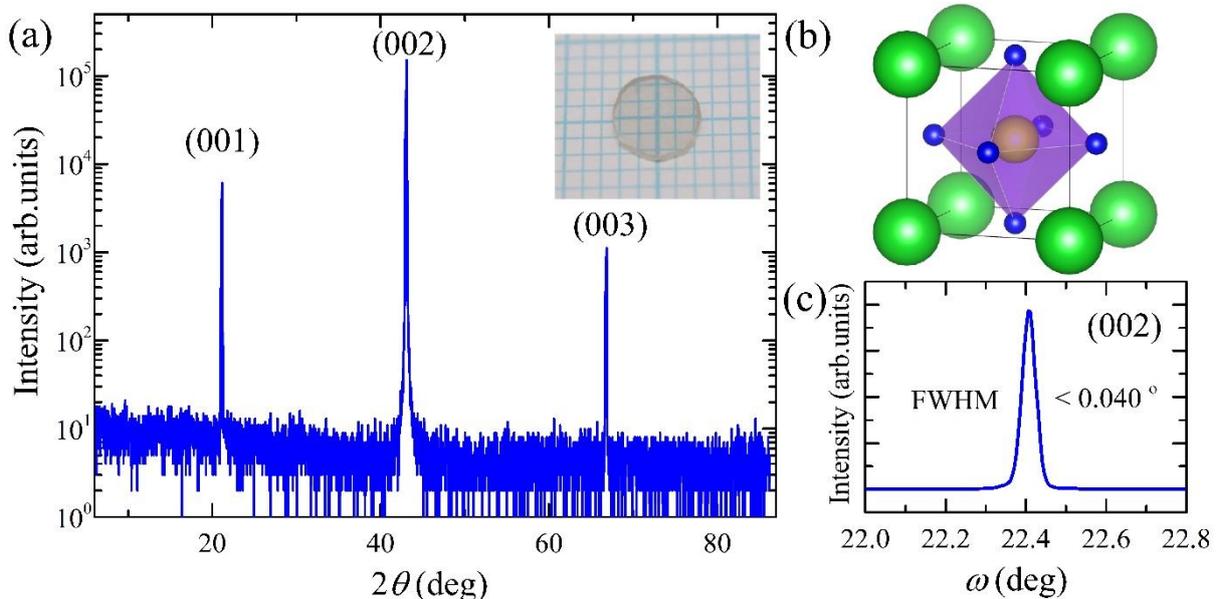

**Figure 1.** (**a**) The $\theta$-$2\theta$ scan of x-ray diffraction, showing (00*l*) (*l* =1,2, and 3) peaks of the BaZrO$_3$ single crystal. The inset image shows the polished (001) surface of a BaZrO$_3$ single crystal on a graph paper with each line spacing of 1 mm. (**b**) The cubic perovskite structure of BaZrO$_3$, in which Ba (green spheres), Zr (yellow sphere) and oxygen ions (blue spheres) are coordinated. (**c**) The rocking curve ($\omega$-scan) measured at the (002) peak of the BaZrO$_3$ single crystal, exhibiting the full-width-half-maximum (FWHM) of less than 0.040 °.

*2.3. Infrared spectroscopy experiment at the atmospheric pressure*

In order to identify the exact energies of zone-center optical phonons at atmospheric pressure, IR reflectivity of the (001) plane of the BZO crystal was measured from 100 to 8000 cm$^{-1}$ by using Fourier-transform infrared spectrometer with in-situ gold overcoating technique. The real and imaginary parts of dielectric function were obtained by Kramers-Kronig transformation of the reflectivity data. For the transformation, complex dielectric function in the energy between 6000 and 50000 cm$^{-1}$ obtained by spectroscopic ellipsometer was used. The reflectivity below the low-frequency cutoff of our measurements was extrapolated as a constant.

## 3. Results and Discussions

*3.1. Vibrational modes at ambient pressure*

Figure 2(a) and (b) show polarized Raman spectra of a polycrystalline BaZrO$_3$ ceramic specimen and a BaZrO$_3$ single crystal, respectively. The overall features in the Raman spectra of the polycrystal qualitatively agree with that of the single crystal. Moreover, the Raman data in Fig. 2(b) nearly reproduce recently reported Raman scattering results on the single crystals at ambient pressure [16,31]. However, it should be noted that several additional peaks, located at 179, 248, 392, and 473 cm$^{-1}$, are only found in the ceramic sample [Fig. 2(a)] but are not clearly identified in the single crystal [Fig. 2(b)]. In the frequency window of those additional peaks, the single crystal data merely exhibit a hump or a broadened peak feature. This indicates that the additional peaks are broadened or suppressed in the single crystal. It is further noted that a larger number of peaks have been found in the previously reported Raman spectra of the BZO ceramics [33,34]. The previous and current Raman spectra on the ceramic specimens thus indicate that scattering amplitudes of several additional phonons might be enhanced in the polycrystal, presumably due to the presence of local distortion or disorder.

It is well known that the cubic perovskite with the space group $Pm\bar{3}m$ allows three pairs of IR-active transverse optical (TO) and longitudinal optical (LO) phonon modes with the irreducible representation (irrep) $\Gamma_4^-$, and one degenerate IR- and Raman- inactive mode, called a silent mode, with irrep $\Gamma_5^-$. Because the three IR-active TO and LO phonons are only IR-active, the phonon frequencies as determined from the IR measurements can be useful for identifying phonons observed in the Raman spectra. Although two former measurements on the IR phonon spectra of BZO ceramics have been reported [34,35], their phonon frequencies exhibit sizable discrepancies up to 110 cm$^{-1}$. Therefore, we have measured the IR reflectivity of the BZO single crystal and obtained the dielectric function after the Kramers-Kronig transformation as shown in Figs. 2(c) and (d). To identify the TO and LO frequencies, we have chosen peak and zero positions of imaginary and real part of dielectric functions, respectively. Thus obtained frequencies of each TO and LO phonon are summarized in Table 1, which shows a good agreement with the latest first-principles calculations results [24,25]. Table 1 also summarizes the Raman peaks found in both ceramic [Fig. 2(a)] and single crystal [Fig. 2(b)] specimens.

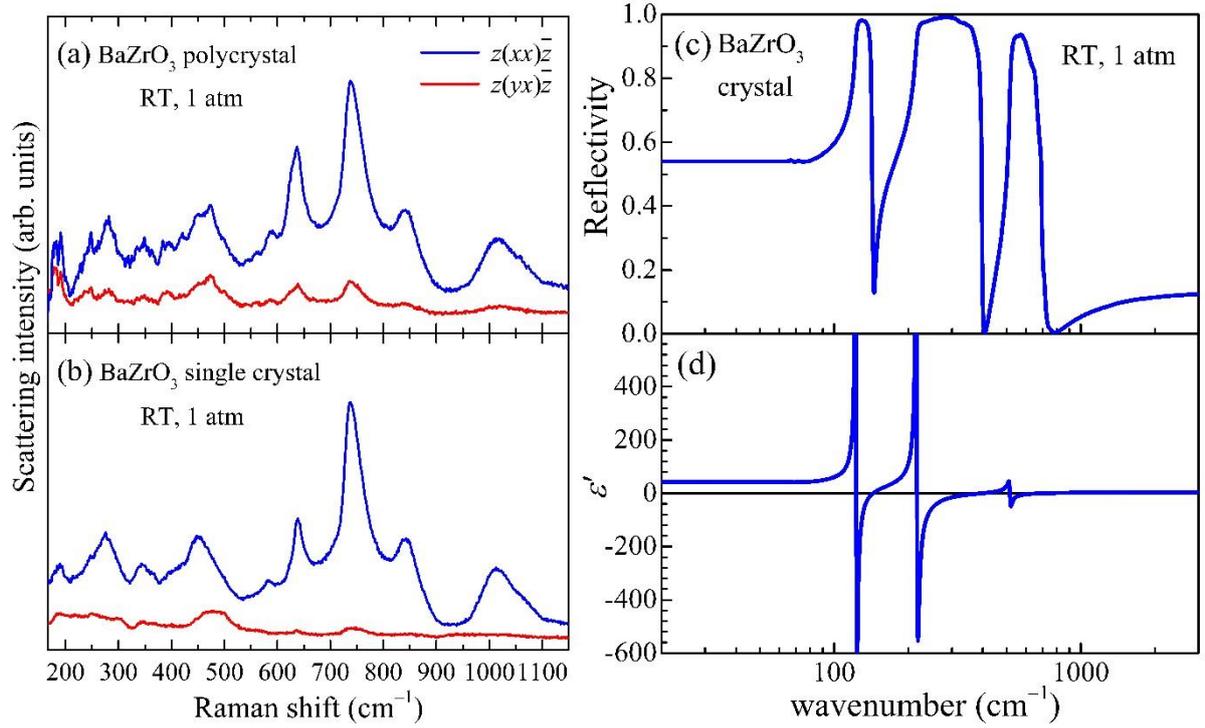

**Figure 2.** Vibrational spectra of BaZrO3 measured at 1 atm. (**a**) Raman spectra of the BaZrO3 polycrystal. (**b**) Raman spectra of the BaZrO3 single crystal. (**c**) Infrared reflectivity spectrum of the BaZrO3 single crystal. (**d**) Real part of dielectric function $\varepsilon'$ of the BaZrO3 single crystal obtained from the Kramers-Kronig transformation of the reflectivity data in (c).

As evident from direct comparison with the IR data, the main Raman peak positions do not coincide with those zone-center $\Gamma_4^-$ phonons frequencies, i.e., TO$_i$ and LO$_i$ ($i$=1-3) mode frequencies observed in the IR spectra. The fact that each $\Gamma_4^-$ phonon remains Raman-inactive rules out the 45 crystal structures involving polar $\Gamma_4^-$ displacements among descendent 60 structures of the parental $Pm\bar{3}m$ lattice [36]. In addition, we confirmed that respective zone-boundary phonons at $R$, $M$, or $X$ points with each energy predicted by the DFT computations [25] even fails to match with the measured Raman signals (as tabulated in Table 1). If the symmetry were lowered by distortion, at least one of phonons located at the relevant $k$-points of cubic perovskites ought to become a Raman-active mode by relocating itself into the new Γ point in a folded zone of the supercell. For example, AFD $R_4^+$ distortions can construct one of three symmetries, tetragonal $I4/mcm$, orthorhombic $Imma$, or rhombohedral $R\bar{3}c$ structures, all of which can make $R$ point phonons become Raman-active while leaving the original Γ point phonons Raman-inactive. However, it is found that major peak frequencies in the measured Raman data disagree with those predicted by $R$ or $M$ point phonons (Table 1), clearly precluding the possibility of other 15 lattice structures induced by AFD octahedra rotations with irreps $M_3^+$ and $R_4^+$ [37].

To fully understand the Raman spectra of the BZO crystal, the multi-phonon excitations should be then considered. In a Raman scattering process, the crystal with inversion symmetry indeed allows the creation of phonon pairs throughout the Brillouin zone (BZ) with opposite wave vectors, thus satisfying the momentum conservation [38]. It is customary to regard combinations and overtones of phonons at the k-points with high symmetries such as $\Gamma$, $R$, $M$, and $X$ of cubic perovskites since the scattering rates of two-phonon modes are weighted by the phonon density of states. It is indeed found that the two-phonon energies at $\Gamma$, $R$, and $M$ points are sufficient to assign the observed modes completely as summarized in Table 1; one can corroborate that $\Gamma_4^-$ phonons frequencies as determined by the IR spectroscopy comprise the majority of the assignment such as $2LO_1$, $2TO_2$, $TO_1+TO_2$, $TO_2+TO_3$, and so on. On the other hand, solo phonons located at the zone center and boundaries, implying the short-ranged or local lattice distortions [39,40], are unnecessary to explain the overall frequencies in the spectra. Therefore, the Raman scattering data at the atmospheric pressure can be unambiguously identified by the multi-phonon excitations, consequently supporting the cubic $Pm\bar{3}m$ symmetry as a structural ground state of the BZO crystal.

Table 1. A summary of the BaZrO$_3$ optical phonons at 1 atm. The first and second columns denote the observed Raman shifts, the third column lists the frequencies of the zone-center phonons determined from the IR spectra, and the fourth column shows the calculated frequencies of the relevant zone-boundary phonons in the literatures. Frequencies are presented in wavenumbers (cm$^{-1}$). Abbreviations in the fifth column are as follows: out-of-phase antiferrodistortive mode (OOP-AFD), A-type antiferroelectric mode (A-AFE), in-phases antiferrodistortive mode (IP-OOP), G-type antiferroelectric mode (G-AFE), scissor mode (Scissor), and Jahn-Teller-like rotation mode (JT-Rot).

| Polycrystal Raman shift [Fig. 2(a)] | Single crystal Raman shift [Fig. 2(b)] | Single crystal IR mode [Fig. 2(d)] | Zone-boundary phonon (DFT) [Refs. 24 and 25] | Assignment |
|---|---|---|---|---|
| 179 | – | | | $2X_5^+$ (Ba) |
| 199 | 189 | | | $2M_3^+$ (O$_6$) |
| 248 | – | | | $2TO_1$ |
| 280 | 276 | | | $2LO_1$ |
| 348 | 343 | | | $TO_1+TO_2$ |
| 392 | – | | | $LO_2$ |
| 444 | 447 | | | $2TO_2$ |
| 473 | – | | | $R_5^+ + R_5^+$ (Ba,O$_6$) |
| 588 | 584 | | | $R_4^+ + R_3^+$ (O$_6$) |
| 637 | 639 | | | $TO_1+TO_3$ |
| 738 | 738 | | | $TO_2+TO_3$ |
| 842 | 843 | | | $LO_1+LO_3$ |
| 1017 | 1013 | | | $2TO_3$ |
| | | 123 | | $TO_1$ (Ba–O) |
| | | 144 | | $LO_1$ (Ba–O) |
| | | 217 | | $TO_2$ (Zr–O) |
| | | 398 | | $LO_2$ (Zr–O) |
| | | 516 | | $TO_3$ (O–O) |
| | | 701 | | $LO_3$ (O–O) |
| | | | 48 | $R_4^+$ (O$_6$, OOP-AFD) |
| | | | 87 | $X_5^+$ (Ba, A-AFE) |
| | | | 95 | $M_3^+$ (O$_6$, IP-AFD) |
| | | | 106 | $R_5^+$ (Ba, G-AFE) |
| | | | 373 | $R_5^+$ (O$_6$, Scissor) |
| | | | 545 | $R_3^+$ (JT-Rot.) |

*3.2. Raman scattering at higher pressures*

Raman spectra of the BaZrO$_3$ single crystals measured at high pressures are displayed in Figure 3. It turns out that the overall Raman modes featured in the high-pressure spectra are analogous to those in Ref. 19, in which Raman modes of BaZrO$_3$ ceramics were reported at high pressures. It is important to notice in Figure 3 that several new peaks start to appear from 8.4 GPa, and one of them at ~390 cm$^{-1}$ splits into two above 19.2 GPa [Fig. 3(c)]. All these findings indicate that there exist two major structural changes at ~8.4 and ~19.2 GPa. The mode frequencies as determined from Lorentzian fits are summarized in Fig. 3 at each pressure. The symbols used in the legend indices in Fig. 4(a) represent Raman frequencies obtained from the Lorentzian fits in the phonon spectra shown in Fig. 3(a) and (b); note that the same symbols are used to represent the corresponding phonon modes in Figs. 3 (a) and (b). It is found that all the newly observed peaks above ~8.4 GPa can be assigned as the phonons stemming from the *R* point of the original cubic BZ (*vide infra*).

One of the key observations in the high-pressure spectra is the presence of two peaks ≈ 390 cm$^{-1}$ above 19.2 GPa [Figs. 3 and 4(a)]. It is a single peak in the pressure range between 8.4 and 18.6 GPa, whose extrapolation down to 0 GPa arrives at a frequency around 370 cm$^{-1}$. This frequency corresponds to a Raman inactive, zone-boundary $R_5^+$ phonon, which describes the scissor mode of the octahedra in the $Pm\bar{3}m$ phase, according to the DFT calculations [24,25]. According to the symmetry analysis, this mode is supposed to split into two Raman-active modes (E$_g$ and B$_{2g}$) in the tetragonal *I4/mcm* structure, three Raman-active modes (B$_{2g}$, B$_{3g}$, and A$_g$) in the orthorhombic *Imma* structure, and one Raman-active (E$_g$) and another Raman inactive (A$_{2g}$) modes in the rhombohedral $R\bar{3}c$ structure. Therefore, the appearance of a single peak above 8.4 GPa and its splitting above 19.2 GPa are again supportive of the two successive structural phase transitions at each pressure, i.e, one from the cubic to the rhombohedral structure and another from the rhombohedral to the tetragonal structure.

Among the two separated modes, the strength of the B$_{2g}$ peak should be suppressed in the parallel-polarized spectra by symmetry. Despite the finite leakage from the diamond anvils, the relative amplitude of the peak with lower energies (marked with orange lozenges) is clearly reduced in the parallel-polarized spectra [Fig. 3(a)]. Therefore, we could further assign the different symmetries of the two peaks; one with lower frequency can be assigned to the B$_{2g}$ and the other to the E$_g$ modes. The splitting of the peak into two above 19.2 GPa agrees well with the data of polycrystals [19]. Nonetheless, the authors in Ref. 19 interpreted the peak splitting as a signature of a phase transition from the $R\bar{3}c$ to the *Imma*, which requires three split peaks above 19.2 GPa. We believe this assignment is inconsistent with the experimental finding of the two peaks above 19.2 GPa.

Emergence of other peaks are also consistent with the multiple phase transitions. The high frequency peak located ~630 cm$^{-1}$ starts to appear above ~14 GPa and increases its frequency with increasing pressure. Therefore, its Raman frequency is extrapolated to ~580 cm$^{-1}$ upon being extended into 0 GPa as indicated by a violet dashed line in Fig. 4(a). The mode frequency of ~560 cm$^{-1}$ at 0 GPa is roughly close to the $R_3^+$ mode frequency of 545 cm$^{-1}$ (Table 1) in the cubic symmetry, which describes the Jahn–Teller-like rotation mode of the octahedra according to the DFT calculations [24,25]. The Jahn–Teller-like rotation mode generates one Raman-active E$_g$ mode in the $R\bar{3}c$ phase and one Raman-active B$_{1g}$ mode in the tetragonal phase, being consistent with the observed data. The peak abruptly gains intensity above 19.2 GPa in the unpolarized spectra [Fig. 3(b)] while it remains silent in the parallel-polarized spectra [Fig. 3(a)]. As the B$_{1g}$ mode is generally known to appear in the cross-polarization spectra, the strongly enhanced intensity above 19.2 GPa clearly supports the assignment of the phonon mode to the B$_{1g}$ symmetry in the tetragonal phase. If the system were to be in the orthorhombic *Imma* symmetry, this mode would result in two Raman-active modes. Therefore, the evolution of the E$_g$ mode in the $R\bar{3}c$ phase into only one B$_{1g}$ mode above 19.2 GPa again supports that the structural phase transition at 19.2 GPa should be from the rhombohedral to the tetragonal structure.

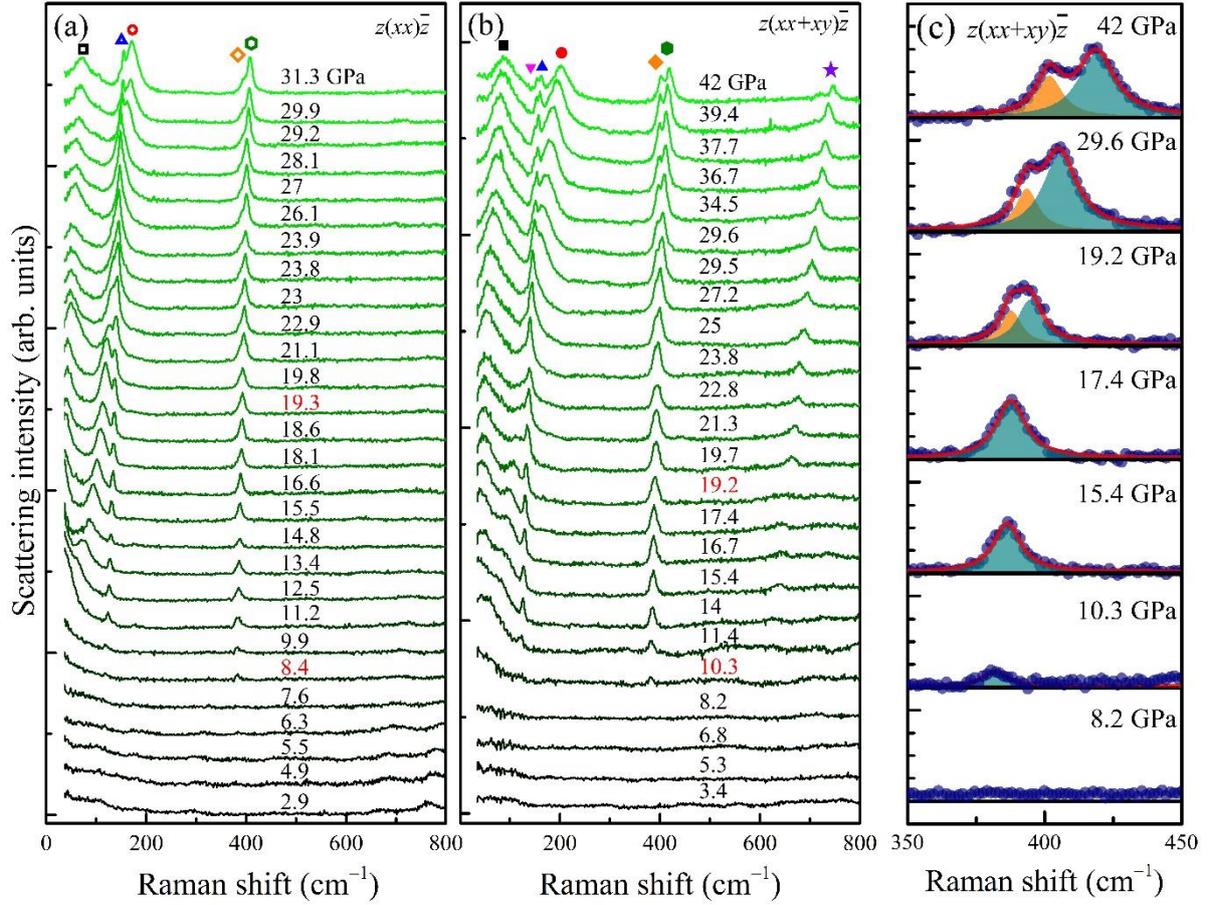

**Figure 3.** Pressure-dependent Raman scattering of the BaZrO$_3$ single crystal. (**a**) $z(xx)\bar{z}$ spectra measured with the polarizing filter. (**b**) $z(xx+xy)\bar{z}$ spectra measured without the polarizing filter. The Raman phonon modes are indicated as the symbols in (a) and (b), of which frequencies obtained from the fitting process are summarized in Fig. 4(a) with the same type of symbols. The pressures at which a structural phase transition occurs are colored in red. (**c**) The evolution of the phonon mode, stemming from an $R_5^+$ scissor mode (Raman inactive in the cubic phase), at a few representative pressures. Navy dots exhibit measured data. The Lorentzian fits shown as pale green and orange areas represent the E$_g$ and B$_{2g}$ modes, respectively, and the red curves show a sum of the two modelled Lorentzian contributions.

Another peak located at 120 cm$^{-1}$ develops in the $R\bar{3}c$ phase above 8.4 GPa and extrapolates to the frequency ~110 cm$^{-1}$ at 0 GPa, which is close to the $R_5^+$ mode frequency of 106 cm$^{-1}$ (Table 1) as predicted by the DFT calculations for describing the antiphase antiferroelectric (AFE) motions of Ba atoms [24,25]. The observation of one mode is consistent with the prediction that there exists one Raman-active mode (E$_g$) and one Raman-inactive mode (A$_{2g}$) in the $R\bar{3}c$ phase. At pressures above 19.2 GPa in the tetragonal *I4/mcm* structure, it is expected to present two Raman-active modes (E$_g$ and B$_{2g}$). Remarkably, we are able to resolve the two peaks above 30 GPa out of the considerable overlap of the two

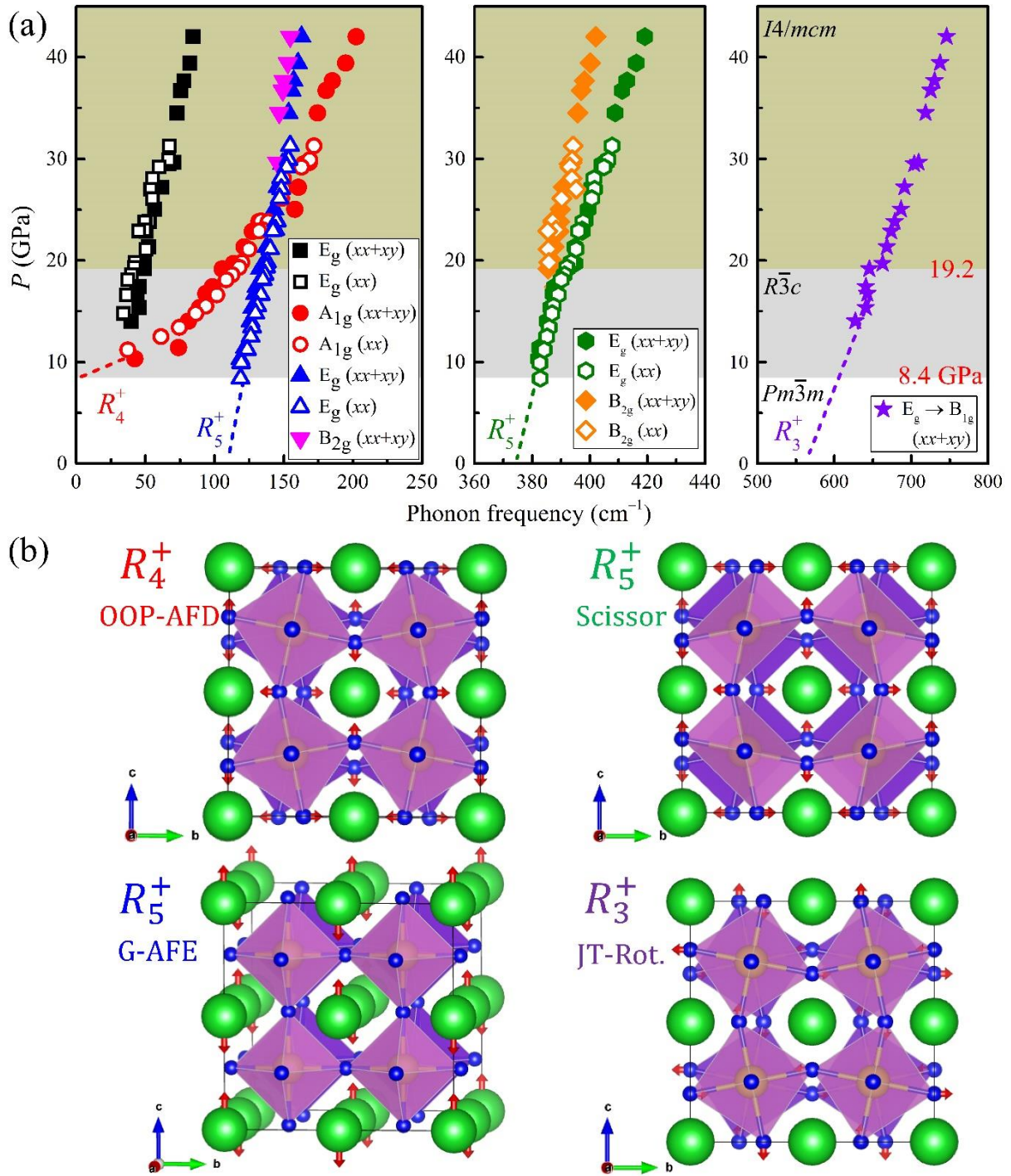

**Figure 4.** (**a**) Evolution of the Raman mode frequencies with variation of pressure. Open and solid marks are obtained from the Figs. 3(a) and 3(b), respectively. Dashed lines represent putative extension of each mode to the zero pressure. (**b**) Atomic displacements of *R* point phonons illustrated on the 2×2×2 supercell with red arrows. Note that the lattice deformations are exaggerated. Here, the acronyms refer to out-of-phase antiferrodistortive mode (OOP-AFD), G-type antiferroelectric mode (G-AFE), scissor mode (Scissor), and Jahn-Teller-like rotation mode (JT-Rot).

modes (inverted magenta and upright blue triangles in Figs. 3(a),(b) and 4). Furthermore, the higher-frequency peak (upright blue triangles) survives in the parallel-polarized spectra

[Fig. 3(a)], which is expected for the $E_g$ mode. Accordingly, the peak with a lower frequency (inverted magenta triangles) is assigned as the $B_{2g}$ mode.

At the lowest measured frequencies, two peaks located ~ 43 cm$^{-1}$ appear above 10.3 GPa, which can be assigned as the out-of-phase AFD $R_4^+$ soft mode [24,25]. In the cubic symmetry at 0 GPa, this AFD mode ($R_4^+$) should be close to ~48 cm$^{-1}$ according to the DFT calculations (Table 1). However, due to the soft mode feature, the $R_4^+$ mode frequency near the structural phase boundary of 8.4 GPa is extrapolated to become zero. At higher pressures above 8.4 GPa, this AFD mode seems to be gradually splitting into doubly degenerate $E_g$ and nondegenerate $A_{1g}$ modes, both of which are stabilized in both $R\bar{3}c$ and $I4/mcm$ phases. The two corresponding modes can be clearly distinguishable above 11.4 GPa, and their frequencies become further apart with the pressure increase. The relative intensity of the higher-frequency mode (red circles) is pronounced in the parallel-polarized spectra corresponding to the $A_{1g}$ irrep, while the lower-frequency $E_g$ peak (black squares) exhibits robust relative intensity in both polarized and unpolarized spectra as compared to other $E_g$ modes.

In the assignment of the crystal symmetry based on the Raman spectra, there is always a possibility that the actual crystal symmetry can be lower than the assignment as the expected Raman active modes in a given structure may not be fully resolved in the experiments. Therefore, to validate the assigned crystal symmetry in the high pressure range, it is worthwhile to compare at least the number of measured phonons with the expected ones based on the symmetries. It is known in the rhombohedral $R\bar{3}c$ perovskites that among the total 18 optical phonons ($A_{1g}$ + 2$A_{1u}$ + 3$A_{2g}$ + 3$A_{2u}$ + 5$E_u$ + 4$E_g$), only 5 optical phonons ($A_{1g}$ + 4$E_g$) can appear in the Raman spectra. Besides, in the tetragonal $I4/mcm$ perovskites, among the total 19 optical phonons ($A_{1g}$ + $A_{1u}$ + 2$A_{2g}$ + 3$A_{2u}$ + $B_{1g}$ + $B_{1u}$ + 2$B_{2g}$ + 5$E_u$ + 3$E_g$), only 7 optical phonons ($A_{1g}$ + $B_{1g}$ + 2$B_{2g}$ + 3$E_g$) are known to become Raman-active. As summarized in Fig. 4(a), all the expected number of phonons for the assigned crystal symmetry are indeed confirmed, namely five modes above 8.4 GPa and seven modes above 19.2 GPa. Consequently, the observed high-pressure Raman modes coherently support the two-step structural transition from the cubic to rhombohedral to tetragonal phases, and all the irreps of the modes can be successfully assigned according to each lattice structure.

It is noted that the pressure-dependent measurements presented in Figs. 3(a) and (b), albeit independent two runs, have exhibited nearly consistent phonon mode behaviors, as confirmed in the open and solid symbols in Fig. 4(a). On the other hand, given that the 4:1 methanol-ethanol liquid used as the pressure medium freezes and starts to lose hydrostaticity from 10 GPa, it might be still necessary to confirm the experimental observations by using a different pressure medium offering more reliable hydrostatic conditions. Hence, the pressure-dependent Raman experiment has been repeated using the NaCl as a pressure medium, which is known to provide decent quasi-hydrostaticity up to a pressure above 20 GPa [41], comparable to inert gases such as $N_2$ and Ar [42]. As summarized in Figure S1 of Supplementary Materials, the first structural transition is found at 8.9 GPa, and the second transition occurs at a pressure above 18.6 GPa when NaCl is used. In addition, the observed phonon evolutions in the two pressure ranges (8.9 ≤ $P$ ≤ 18.6 GPa, and 18.6 GPa < $P$) are consistent with the results measured with the 4:1 methanol-ethanol mixture in Figs. 3 and 4. Based on these results, we conclude that the two pressure-induced phase transitions should be understood as the inherent nature of BZO crystals independent of the choice of pressure medium.

It is worthwhile to discuss how the structure transitions can occur in the BZO by application of pressure. Firstly, the large ionic radius of Ba as well as the smaller ionic charge of Ba$^{2+}$ than Zr$^{4+}$ ensures that the compressibility of the BaO$_{12}$ polyhedra becomes greater than that of the ZrO$_6$ octahedra, causing the ZrO$_6$ octahedra to tilt under pressure [43]. Secondly, if the octahedra rotations bring the oxygen atoms and their next-nearest-neighbor Zr atoms closer, the hybridization between the empty Zr-3$d$ and O-2$p$ states renders the next-nearest-neighbor Zr–O interactions become stronger to drive the AFD distortion [43]. Finally, continuous lattice deformation by applied pressures should yield elastic strain that renormalizes the structural anisotropy from [111]$_{pc}$ to [001]$_{pc}$ [26]. Therefore, the energetic stability of the $R\bar{3}c$ and $I4/mcm$ phases can be switched by compression, in agreement with the observed second transition at 19.2 GPa.

A recent Brillouin scattering experiment [21] has claimed that the inversion symmetry breaking may occur in BZO at room temperature and at ambient pressure in a form of short-ranged ferroelectric (FE) distortions. If the claim is true, one may argue that the intermediate $R\bar{3}c$ state may be helpful to develop the short-range FE distortion near the Ba site by stabilizing the noncentrosymmetric $R3c$ phase at ambient pressure. However, the FE mode which takes the $R\bar{3}c$ structure to the noncentrosymmetric $R3c$ phase should involve considerable Ba displacements, which is unlikely to occur for the large Ba$^{2+}$ ions [44]. Furthermore, if the short-range FE distortion occurs at the Zr sites, it is expected that the phonon modes found in the IR spectra should also become Raman active. However, no sign of the first order Raman active phonon mode is found at ambient pressure, supporting that inversion symmetry is preserved in the high-quality BZO single crystal synthesized by optical floating-zone technique.

## 4. Conclusions

In conclusion, the structure of BaZrO$_3$ single crystals has been determined at atmospheric and high pressures by the Raman scattering measurements. All the Raman modes observed at ambient pressure can be assigned to the multi-phonon scatterings allowed in the cubic perovskite structure by referring to the complementary infrared phonon data and recent DFT calculations. Based on the observation of the appearance of new peaks at high pressures and their splitting, the two pressure-induced structural phase transitions are identified, being $Pm\bar{3}m$ → $R\bar{3}c$ at 8.4 GPa and $R\bar{3}c$ → $I4/mcm$ at 19.2 GPa. The evolution of the spectra at high pressures undoubtedly originates from the octahedra rotations; therefore, the absence of such features at atmospheric pressure corroborates the cubic $Pm\bar{3}m$ structure as the ground state of BaZrO$_3$ near 0 GPa. Moreover, the results imply that the $R\bar{3}c$ structure is closer to the ground state than the $I4/mcm$ or $Imma$ phases. It is expected that the comprehensive understanding on the intrinsic structural phases of the BZO crystal could be helpful for strain engineering or chemical substitution, as well as the characterizations of other crystalline forms such as ceramics and nanocrystals relevant to industrial applications.

**Supplementary Materials:** The following is available online; Figure S1: High-pressure evolution of BaZrO$_3$ phonons measured with NaCl pressure medium.

**Author contributions:** Conceptualization, D.-H.G. and K.H.K.; methodology, Y.S., Y.H.L., J.H.L., S.M. and Y.S.O.; validation, D.-H.G., S.M. and Y.S.O.; formal analysis, D.-H.G., S.M. and Y.S.O.; investigation, D.-H.G., S.M. and Y.S.O.; resources, Y.S.O.; data curation, D.-H.G., S.M. and Y.S.O.; writing—original draft preparation, D.-H.G.; writing—review and editing, S.M., Y.S.O. and K.H.K.; visualization, D.-H.G. and Y.S.O.; supervision, K.H.K.; project administration, K.H.K.; funding acquisition, S.M., Y.S.O. and K.H.K. All authors have read and agreed to the published version of the manuscript.

**Funding:** This research has been financially supported by the National Research Foundation of Korea (NRF) grant funded by the Korean government (NRF-2019R1A2C2090648), Korea Basic Science Institute (National Research Facilities and Equipment Center) grant funded by the Ministry of Education (Grant No. 2021R1A6C101B418), and Industry-University collaboration grant (0409-20200269) by Samsung Electronics. The work at HYU has been supported by the National Research Foundation grant funded by the Korean government (MSIT) (2022R1F1A1072865). Y.S.O. acknowledges support from the Basic Science Research Programs through the National Research Foundation of Korea (2020R1A2C1009537).

**Conflicts of Interest:** The authors declare no conflict of interest.